\documentclass[acmsmall]{acmart}

\usepackage{tikz}
\usepackage{xspace}
\usepackage{enumitem}
\usepackage{filecontents}
\usepackage{pifont}
\usepackage{subfig}
\usepackage{diagbox}
\usepackage{pgfplots}
\usepackage{float}
\usepackage{color}
\usepackage{tablefootnote}
\usepackage{multirow}
\usepackage{mathrsfs,amsmath}
\usepackage{makecell}
\usepackage{bm}
\usepackage{color, colortbl}
\usepackage{threeparttable}
\usepackage{pgfplots}

\newcommand{\sysname}{DxPU\xspace}
\newcommand{\proxyname}{DxPU\_PROXY\xspace}
\newcommand{\managername}{DxPU\_MANAGER\xspace}

\newcommand{\tabincell}[2]{\begin{tabular}{@{}#1@{}}#2\end{tabular}}

\AtBeginDocument{%
  }

\begin{document}

\title{DxPU: Large Scale Disaggregated GPU Pools in the Datacenter}

\author{Bowen He}
\orcid{0000-0001-7794-2520}
\affiliation{%
  \institution{Zhejiang University and Alibaba Group}
  \city{HangZhou}
  \country{China}}
\email{hebowen.hbw@alibaba-inc.com}

\author{Xiao Zheng}
\orcid{0000-0003-3191-5636}
\affiliation{%
  \institution{Alibaba Group}
  \city{HangZhou}
  \country{China}}
\email{zhengxiao.zx@alibaba-inc.com}

\author{Yuan Chen}
\orcid{0009-0009-2321-4910}
\affiliation{%
  \institution{Zhejiang University and Alibaba Group}
  \city{HangZhou}
  \country{China}}
\email{chenyuan.chen@alibaba-inc.com}

\author{Weinan Li}
\orcid{0009-0005-6923-5312}
\affiliation{%
  \institution{Alibaba Group}
  \city{HangZhou}
  \country{China}}
\email{william.lwn@alibaba-inc.com}

\author{Yajin Zhou}\thanks{Yajin Zhou is the corresponding author.} 
\orcid{0000-0001-7610-4736}
\affiliation{%
  \institution{Zhejiang University}
  \city{HangZhou}
  \country{China}}
\email{yajin_zhou@zju.edu.cn}

\author{Xin Long}
\orcid{0009-0002-4528-0993}
\affiliation{%
  \institution{Alibaba Group}
  \city{HangZhou}
  \country{China}}
\email{longxin.xl@alibaba-inc.com}

\author{Pengcheng Zhang}
\orcid{0009-0001-6740-227X}
\affiliation{%
  \institution{Alibaba Group}
  \city{HangZhou}
  \country{China}}
\email{binxu.zpc@alibaba-inc.com}

\author{Xiaowei Lu}
\orcid{0009-0009-9230-3589}
\affiliation{%
  \institution{Alibaba Group}
  \city{HangZhou}
  \country{China}}
\email{dege.lxw@alibaba-inc.com}

\author{Linquan Jiang}
\orcid{0009-0004-4624-5174}
\affiliation{%
  \institution{Alibaba Group}
  \city{HangZhou}
  \country{China}}
\email{linquan.jlq@alibaba-inc.com}

\author{Qiang Liu}
\orcid{0009-0006-5792-322X}
\affiliation{%
  \institution{Alibaba Group}
  \city{HangZhou}
  \country{China}}
\email{yf.yifeng@alibaba-inc.com}

\author{Dennis Cai}
\orcid{0009-0001-7272-8143}
\affiliation{%
  \institution{Alibaba Group}
  \city{HangZhou}
  \country{China}}
\email{d.cai@alibaba-inc.com}

\author{Xiantao Zhang}
\orcid{0009-0006-3065-7646}
\affiliation{%
  \institution{Alibaba Group}
  \city{HangZhou}
  \country{China}}
\email{xiantao.zxt@alibaba-inc.com}
\renewcommand{\shortauthors}{B. He et al.}

\begin{abstract}
The rapid adoption of AI and convenience offered by cloud services have resulted in the growing demands for GPUs in the cloud.
Generally, GPUs are physically attached to host servers as PCIe devices.
However, the fixed assembly combination of host servers and GPUs is extremely inefficient in resource utilization, upgrade, and maintenance.
Due to these issues, the GPU disaggregation technique has been proposed to decouple GPUs from host servers. It aggregates GPUs into a pool, and allocates GPU node(s) according to user demands.
However, existing GPU disaggregation systems have flaws in software-hardware compatibility, disaggregation scope, and capacity.

In this paper, we present a new implementation of \textit{datacenter-scale} GPU disaggregation, named \sysname.
\sysname efficiently solves the above problems and can flexibly allocate as many GPU node(s) as users demand.
In order to understand the performance overhead incurred by \sysname, we build up a performance model for AI specific workloads.
With the guidance of modeling results, we develop a prototype system, which has been deployed into the datacenter of a leading cloud provider for a test run.
We also conduct detailed experiments to evaluate the performance overhead caused by our system. The results show that the overhead of \sysname is less than 10\%, compared with native GPU servers, in most of user scenarios.

\end{abstract}




\maketitle

\section{Introduction}

\label{sec:introduction}
The great success of Artificial Intelligence (AI) has boost demands for Graphics Processing Units (GPUs).
In pursuit of \textit{scalability}, \textit{elasticity}, and \textit{cost efficiency},
users tend to rent GPUs in the cloud to accelerate their AI workloads.
Hence, cloud providers leverage \textit{datacenter-scale} GPU clusters to satisfy different user requirements.

For GPU clusters in the cloud, GPUs are attached to the host server
as PCIe devices with fixed proportions (e.g., 2, 4, 6, or 8 GPUs). 
With the help of I/O virtualization~\cite{GPU_virtualization}, CPUs and GPUs can be splitted and allocated to several
virtual machines (VMs) \textit{on the same host server}.
However, the virtualization techniques fail to support flexible GPU provisioning \textit{between host servers}.
For instance, even if one or more GPUs on one host sever are free, they cannot be used by VMs on a different host server.
In this server-centric architecture~\cite{Composable, Resource_disaggregation}, the fixed coupling of host servers and GPUs brings the following practical limitations for cloud providers.

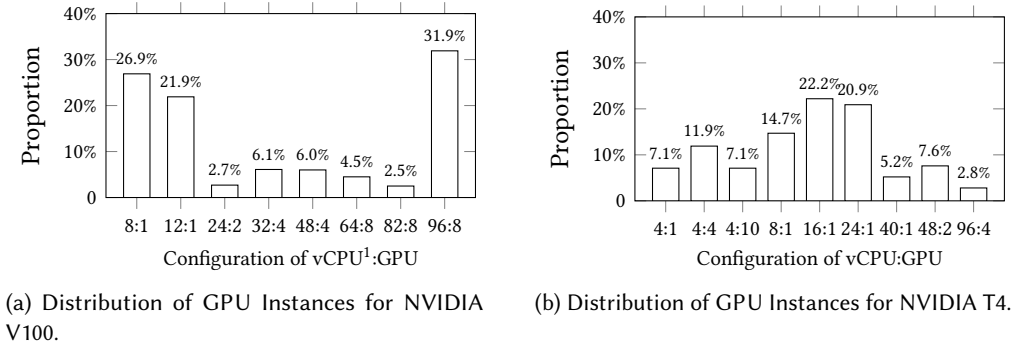
\begin{figure*}[t]
  \subfloat[Distribution of GPU Instances for NVIDIA V100.]{
    \begin{tikzpicture}
      \begin{axis}[ybar,
        xtick={0, 1, 2, 3, 4, 5, 6, 7},
        xticklabels={8:1, 12:1, 24:2, 32:4, 48:4, 64:8, 82:8, 96:8},
        xticklabel style={font=\footnotesize},
        xlabel={Configuration of vCPU\footnotemark :GPU},
        xlabel style={font=\footnotesize},
        ymin=0,
        ymax=40,
        ytick={0, 10, 20, 30, 40},
        yticklabels={0, 10\%, 20\%, 30\%, 40\%},
        ylabel={Proportion},
        ylabel style={font=\footnotesize},
        yticklabel style={font=\footnotesize},
        ylabel near ticks,
        height = 4cm,
        width = 6.5cm,
        nodes near coords,
        nodes near coords style={font=\scriptsize},
        point meta=explicit symbolic
        ]
        \addplot[draw=black, fill=white]
        coordinates {
          (0,26.9) [26.9\%]
          (1,21.9) [21.9\%]
          (2,2.7)  [2.7\%]
          (3,6.1)  [6.1\%]
          (4,6.0)  [6.0\%]
          (5,4.5)  [4.5\%]
          (6,2.5)  [2.5\%]
          (7,31.9) [31.9\%]
        };
      \end{axis}
    \end{tikzpicture}
  }
  \qquad
  \subfloat[Distribution of GPU Instances for NVIDIA T4.]{
    \begin{tikzpicture}
      \begin{axis}[ybar,
        xtick={0, 1, 2, 3, 4, 5, 6, 7, 8},
        xticklabels={4:1, 4:4, 4:10, 8:1, 16:1, 24:1, 40:1, 48:2, 96:4},
        xticklabel style={font=\footnotesize},
        xlabel={Configuration of vCPU:GPU},
        xlabel style={font=\footnotesize},
        ymin=0,
        ymax=40,
        ytick={0, 10, 20, 30, 40},
        yticklabels={0, 10\%, 20\%, 30\%, 40\%},
        ylabel={Proportion},
        ylabel style={font=\footnotesize},
        yticklabel style={font=\footnotesize},
        ylabel near ticks,
        height = 4cm,
        width = 6.5cm,
        nodes near coords,
        nodes near coords style={font=\scriptsize},
        point meta=explicit symbolic
        ]
        \addplot[draw=black, fill=white]
        coordinates {
          (0,7.1) [7.1\%]
          (1,11.9) [11.9\%]
          (2,7.1)  [7.1\%]
          (3,14.7)  [14.7\%]
          (4,22.2)  [22.2\%]
          (5,20.9)  [20.9\%]
          (6,5.2)  [5.2\%]
          (7,7.6) [7.6\%]
          (8,2.8) [2.8\%]
        };
      \end{axis}
    \end{tikzpicture}
  }
  \caption{Distribution of GPU Instances with Different Hardware Configurations in the Cloud.
  User requests are diverse in hardware requirements.}
  \label{fig:distribution_instance}
\end{figure*}

\begin{itemize}[nosep,leftmargin=1em,labelwidth=*,align=left]

    \item CPU and GPU fragments are common in the datacenter.
        We collect the distribution of GPU instances
        with different configurations for NVIDIA V100 and T4 in the datacenter of a leading cloud provider 
        and plot it in Figure \ref{fig:distribution_instance}.
        Obviously, in the cloud, user requests are diverse in resource requirements.
        Considering the diverse and dynamically changing resource requirements of different workloads, 
		  	it is difficult to \textit{statically} equip each host server with the right balance of hardware resources.
        Hence, this causes either CPU or GPU fragments in server clusters and brings additional cost for cloud providers.

    \item The cost of \textit{hardware maintenance} and \textit{upgrade} remains to be reduced.
          On the host server, any single point of failure makes all hardware resources unavailable.
          Also, different hardware (e.g., CPU, GPU, and memory) exhibits distinct upgrade trends~\cite{Resource_disaggregation}.
          Under the fixed coupling scheme, the upgrade of any hardware needs to shutdown the server and influences the availability of other resources.
          As a result, it costs excessive efforts to maintain the \textit{availability} and \textit{up-to-dateness} of the cloud infrastructure at the same time~\cite{Datacenter_cost}.
\end{itemize}

To address these limitations, the \textit{GPU disaggregation}~\cite{EMF} technique has been proposed,
which aims to physically decouple GPUs from host servers and group them into pools.
As a result, GPU workloads from the applications on the host server are forwarded to allocated node(s) in the pool, named as \textit{disaggregated GPU}.
With such a design, GPU provisioning is no longer restricted \textit{on the same host server}.
Confronted with varying requests, cloud providers can dynamically allocate or reclaim  GPUs in the pool without worrying about resource fragments.
Meanwhile, CPUs and GPUs can be managed, scaled, and upgraded independently, optimizing infrastructure expenses.

Throughout the software and hardware stack, we can implement \textit{GPU disaggregation} at the user-mode library, kernel-model driver, and PCIe level.
Yet, systems at the library and driver level have three natural limitations.
First, the applicable software and hardware are restricted.
For example, CUDA-based disaggregation techniques~\cite{bitfusion, rCUDA} cannot be used for applications built on OpenGL.
Second, it is a continuous work for developers to keep the wrapper libraries and drivers up-to-date all the time.
Last, reducing the communication overhead and optimizing the performance are troublesome and time-consuming.
By contrast, exploring forwarding strategies at the PCIe level brings the benefits of \textit{full-scenario support}, \textit{software transparency}, and \textit{good performance} in nature.


However, existing solutions at the PCIe level fall short of the expected \textit{disaggregation scope} and \textit{capacity}.
Specifically, they are all built on the PCIe fabric~\cite{EMF, SmartIO, Liqid}, where servers and switches are connected via PCIe cables.
The scope of disaggregation is limited to \textit{a very small number of racks}, \footnotetext{vCPU is the abbreviation for virtual CPU.}failing to reach the \textit{datacenter scale}~\footnote{There can be hundreds of GPU servers in the datacenter.}.
What's more, to reduce the bit error rate and enable high-speed peer-to-peer communications, the PCIe fabric can only accommodate \textit{dozens of} GPUs, which is far from enough.

In order to solve the above problems,
we propose a new design of \textit{GPU disaggregation} at the PCIe level, named \sysname.
In \sysname, a proxy (\proxyname) is added on both the host server and GPU side.
The proxies are interconnected via a network fabric
and act as the medium for the interaction between the host server and GPUs.
Specifically, \proxyname is responsible for \textit{the conversion between PCIe TLPs and network packets}.
From the host server's perspective, GPUs are still accessed locally over PCIe links.
Thus, \sysname is \textit{transparent} to the user and supports \textit{GPU virtualization} techniques.
Moreover, GPUs are dynamically selected and \textit{hot-plugged into} the host server, meaning they are \textit{dedicated to the host server during use}.
Compared with the PCIe fabric based method, it allows the management of GPU pools on the \textit{datacenter scale}.
In summary, \sysname advances the first step towards exploring \textit{datacenter-scale} \textit{GPU disaggregation} at the PCIe level.

To make such a new design into a ready-to-deploy product in the cloud,
the following question, i.e., \textit{What is the performance impact of \sysname for GPU workloads in the cloud}, needs to be answered \textit{before a prototype is implemented}.
To this end, we introduce a performance model in software.
Specifically, the performance overhead introduced by \sysname mainly comes from the longer latency for the interaction between the host server and GPUs.
We choose the AI scenario, the dominant consumer of GPUs in the cloud, and build up a performance model via API hooking to estimate the performance impact of \sysname.
With the built model, we can demonstrate the performance impact of \sysname for GPU workloads in the cloud.


Guided by the performance model, we build an implementation system of \sysname.
Our evaluation in basic, AI, and graphics rendering scenarios illustrates that the performance of \sysname over the native GPU server, which is better than 90\% in most cases, is able to satisfy cloud providers' requirements.
What's more, the pool can hold up to 512 GPU nodes.
Besides, we point out major factors in the current software stack that undermine the performance,
and explore optimization space of \sysname from the view of software-hardware co-design.
Last but not least, we believe that \sysname can be extended to realize disaggregation of other hardware resources in the cloud, which are instructive for both academia and industry.

In summary, contributions of our study are as follows.

\begin{itemize}[nosep,leftmargin=1em,labelwidth=*,align=left]
	\item We propose a new design of \textit{datacenter-scale GPU disaggregation} system at the PCIe level (Section~\ref{sec:design}), named \sysname.
	It is excellent in \textit{full-scenario support}, \textit{software transparency}, \textit{datacenter-scale disaggregation}, \textit{large capacity}, and \textit{good performance}.
	\item We build up a performance model to estimate the performance impact of \sysname for GPU workloads in the cloud (Section~\ref{sec:performance_model})
	and verify its correctness with the implementation of \sysname (Section~\ref{sec:implementation}).
	\item We make detailed evaluations of \sysname with multiple benchmarks to demonstrate the practicality of it.
	The performance of \sysname over the native one is better than 90\% in most of cases, which satisfies cloud providers' requirements (Section~\ref{sec:evaluation}).
	Moreover, we present the insights to optimize \sysname from the software-hardware co-design perspective (Section~\ref{sec:discussion}).

\end{itemize}

\section{Background And Motivation}

\begin{table*}[t]
    \centering
    \begin{threeparttable}
    \caption{Comparison of Existing GPU Disaggregation Systems at Different Levels.}
    \label{tab:comparison}
  \scriptsize
    \begin{tabular}[b]{cccccc}
        \hline
        \tabincell{c}{\textbf{Implementation} \\ \textbf{Level}}&\textbf{Representatives}&\tabincell{c}{\textbf{Software-Hardware} \\ \textbf{Compatibility}}&\tabincell{c}{\textbf{Scope of} \\
         \textbf{Disaggregation}}&\tabincell{c}{\textbf{Capacity of} \\ \textbf{the GPU Pool}}&\tabincell{c}{\textbf{Overall} \\ \textbf{Performance}\tnote{3}}\\
        \hline
        User-mode Library&rCUDA~\cite{rCUDA}, bitfusion~\cite{bitfusion}& Limited & Datacenter-scale\tnote{1} & Large & 80\%-95\% \\
        \hline
        Kernel-mode Driver&cGPU~\cite{cGPU}& Limited & Datacenter-scale & Large & 80\%-99\% \\
        \hline
        \multirow{2}*{PCIe}&Liqid AI Platform~\cite{Liqid}&Unlimited & Rack-scale\tnote{2} & Small & 99\%-100\% \\
        \cline{2-6}
        ~&\sysname (our solution)& Unlimited & Datacenter-scale & Large & mostly 90\%-99\% \\
        \hline
    \end{tabular}
    \begin{tablenotes}
        \item[1] Datacenter-scale means GPU nodes can be allocated to host servers from anywhere in the datacenter.
        \item[2] Rack-scale means host servers and \textit{disaggregated GPUs} must reside in a very small number of racks.
        \item[3] Performance denotes \textit{GPU compute capability} in the specific implementation over the native one. 
    \end{tablenotes}
    \end{threeparttable}
  \end{table*}

\subsection{Resource Disaggregation}
Traditionally, all hardware resources (e.g., CPU, GPU, and memory) are tightly coupled with host servers.
However, as different hardware exhibits diverse iteration cycles, this server-centric architecture makes it hard to upgrade hardware resources independently.
Recently, there emerges a paradigm shift towards \textit{resource disaggregation} in the datacenter~\cite{intelRack, composableCS, dredbox}.
By decoupling these resources and grouping them into corresponding pools, it is easier to maintain, upgrade them and deploy cutting-edge hardware technologies in time.
For example, based on the abstraction of a Transport layer, NVMe over Fabrics~\cite{NVMeoF} (NVMe-oF) enables memory operations over various kinds of interconnects (e.g., Fibre Channel, RDMA, or TCP).
With the help of NVMe-oF, the storage resource can be maintained, and upgraded independently.

\subsection{GPU Disaggregation}
\label{sec:simulation}
As mentioned aboved, the tight coupling of host servers and GPUs causes resource fragments and restricts the independent maintenance, upgrade of hardware resources.
To overcome these problems, \textit{GPU disaggregation} is proposed.
Specifically, GPU workloads on the host server are encapsulated and redirected to \textit{disaggregated GPUs}.
For example, DGSF~\cite{dgsf} makes use of API remoting to realize \textit{GPU disaggregation} for the severless environment and supports live migration at the same time.
In general, we can implement \textit{GPU disaggregation} at the software level (user-mode library, kernel-model driver), and hardware level (PCIe level).
As the software solutions are based on API hooking, the applicable scenarios and GPU performance are greatly affected.
By contrast, exploring forwarding strategies at the PCIe level can overcome these problems easily.
However, existing solutions at the PCIe level fall short of the expected \textit{disaggregation scope} and \textit{capacity}.
In the following paragraphs, we will first explain why we implement \textit{GPU disaggregation} at the PCIe level (Section~\ref{sec:why_PCIe}).
Then we summarize why existing solutions at the PCIe level fall short (Section~\ref{sec:existing_PCIe}),
which motivates the design of \sysname.
Table~\ref{tab:comparison} makes a comprenhensive comparison between existing \textit{GPU disaggregation} systems.

\subsubsection{Limitations of Solutions at the Software Level}
\label{sec:why_PCIe}
As NVMe~\cite{NVMe} specifies how software interacts with non-volatile memory, it is not difficult to extract a Transport Layer to permit communication between host servers and memory over diverse interconnects.
However, as there doesn't exist a uniform specification of the GPU architecture and driver, it is impossible to build such a Transport Layer for \textit{GPU disaggregation}.
So we can only analyze the invoking sequences throughout the software and hardware stack to determine the implementation levels.
Yet, \textit{GPU disaggregation} systems at the library and driver level have its own limitations which cannot be bypassed.

\noindent
\textbf{Limited Application Scenarios.}
For example, rCUDA~\cite{rCUDA} and VMware bitfusion~\cite{bitfusion} implement \textit{GPU disaggregation} by replacing CUDA library with a wrapper one, meaning confined scenario applicability.
In other words, they cannot be used for other applications (e.g., OpenGL-based applications).
In addition, as CUDA is developed by NVIDIA, users fail to utilize AMD or Intel GPUs.
In contrast, \textit{GPU disaggregation} implemented at the driver level can fulfill user requests for general-purpose computing and graphics rendering, (e.g., Alibaba cGPU~\cite{cGPU}).
However, as GPU drivers for different models are distinct, not all of them are suitable for GPU workload redirecting, meaning GPU models are still restricted.
By comparison, building \textit{GPU disaggregation} systems at the PCIe level can easily circumvent these compatibility restrictions.

\noindent
\textbf{Sensitivity to Frequent Software Updates.}
To improve the products, software development companies tend to release a new version every few months.
Obviously, it is a continuous task to keep the corresponding wrapper software up-to-date all the time.

\noindent
\textbf{Excessive Efforts in Performance Optimization.}
During workload forwarding, the parameter encapsulation, transfer, and extraction will introduce much latency.
Especially for encapsulation and extraction at the software level, they cost much more time compared with those at the PCIe level.
So it requires developers to be good at communication optimization.
In contrast, the performance of \sysname is better than 90\% in most cases without any optimization, which is a huge advantage.

By comparison, exploring forwarding strategies at the PCIe level, which touches low-level hardware, can overcome these limitations easily.


\subsubsection{Existing Solutions at the PCIe Level Fall Short}
\label{sec:existing_PCIe}
Existing solutions of \textit{GPU disaggregation} at the PCIe level utilize the PCIe fabric as its interconnect~\cite{SmartIO, Liqid, Intel_rack_scale_design, NTB_kernel}.
Specifically, the PCIe fabric is composed of several specialized PCIe switches.
Meanwhile, Host servers and GPUs are directly connected to the fabric via PCIe cables.
Unlike common PCIe switches in PCIe specification~\cite{PCIe_specification},
the specialized ones need to isolate different host servers' PCIe domains and map GPU node(s) to their memory space~\cite{EMF}.


Although \textit{GPU disaggregation} on the basis of the PCIe fabric achieves almost native performance, there also exist two tough problems which cannot be solved easily.

\noindent
\textbf{Limited Scope of Disaggregation.}
In the PCIe fabric, switches and servers are connected through PCIe cables.
As the length of PCIe cables is limited, the scope of disaggregation can be only restricted to a very small number of racks~\cite{Resource_disaggregation}. 
Host servers and GPUs are still coupled with each other on the rack scale.
In other words, it lacks the flexibility to manage, scale, and upgrade CPUs or GPUs independently.
As a result, it fails to support \text{GPU disaggregation} on the \textit{datacenter scale}.

\noindent
\textbf{Limited Capacity of the GPU Pool.}
To enable high-speed communications between different devices, the tolerable error rate in PCIe links is extremely low.
Statistically speaking, the bit error rate (BER) must be less than ${10^{-12}}$~\cite{PCIe_Loss_Budget, Impact_bit_errors, PCIe_Basics}.
However, if the scale of the PCIe fabric is too large, the BER will go beyond the limitation and fault location, isolation will be particularly difficult.
Hence, although delicate circuit design can decrease the BER, there still exists an upper limit on the number of \textit{disaggregated GPUs} in the PCIe fabric.
For example, in Liqid Powered GPU Composable AI Platform~\cite{Liqid}, the maximum number of accommodable GPUs is 24.
Confronted with datacenter-scale user requests, the capacity of the PCIe fabric is far from enough.
Setting up too many GPU pools will lead to the resource fragment problems mentioned in Section~\ref{sec:introduction} again.

Compared with solutions mentioned above, \sysname is excellent in \textit{full-scenario support}, \textit{software transparency}, \textit{datacenter-scale disaggregation}, \textit{large capacity}, and \textit{good performance}.

\section{Design}

\label{sec:design}
To begin with, We present our design goals of \sysname.

\begin{itemize}[nosep,leftmargin=1em,labelwidth=*,align=left]

    \item \textbf{G1: Scope.} 
    The scope of disaggregation shouldn't be restricted to racks.
    On the contrary, in a \textit{datacenter-scale GPU disaggregation} system, GPU nodes can be allocated to host servers from anywhere in the datacenter.

    \item \textbf{G2: Capacity.}
    The fabric should accommodate at least \textit{hundreds of} GPU nodes (e.g., 128, 256, or 512).

    \item \textbf{G3: Performance.}
    \sysname should not incur high performance overhead.
\end{itemize}

In the following paragraphs, we will first introduce the architecture of \sysname briefly (Section~\ref{sec:architecture}).
Next, we describe the design details of GPU boxes and \proxyname{s} (Section~\ref{sec:GPU_box} and ~\ref{sec:proxy}).
Then, we build up a performance model to estimate effects of network latency on GPU compute capability in \sysname (Section~\ref{sec:performance_model}), which can guide the implementation of \sysname.
Last, with the assistance of the performance model, we build the implementation system of \sysname.
The performance of \sysname confirms the accuracy of our model in turn(Section~\ref{sec:implementation}).

\begin{figure*}[t]
	\centering
	\subfloat[Physical Architecture of \sysname.
	GPUs are physically decoupled from host servers and put into separate GPU boxes.
	\label{fig:physical_arch}]{
		\includegraphics[width=0.45\textwidth]{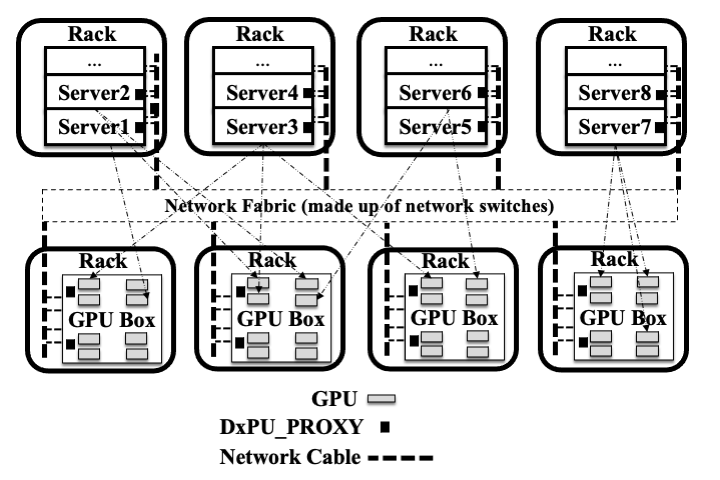}
	}
	\hspace{0.2in}
	\subfloat[Logical Architecture of \sysname.
	\textit{The conversion between PCIe TLP and network packets} is transparent.
    \label{fig:logic_arch}]{
		\includegraphics[width=0.45\textwidth]{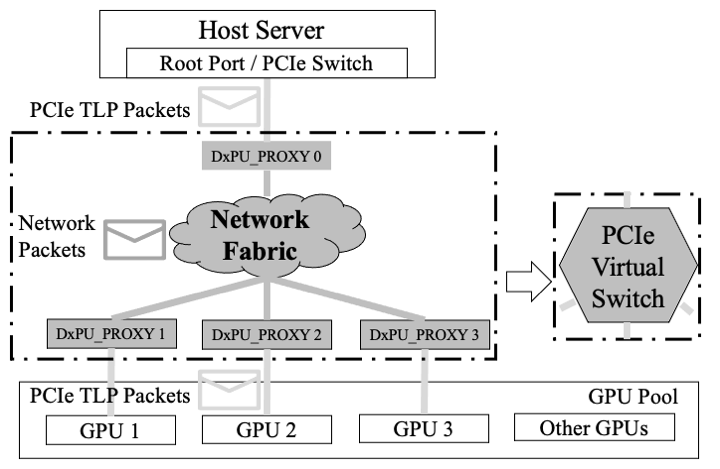}
	}
	\caption{Physical and Logical Architecture of \sysname}
\end{figure*}


\subsection{Architecture}
\label{sec:architecture}
Figure \ref{fig:physical_arch} shows the overall architecture of \sysname.
Compared with the non-disaggregated architecture, which attaches GPUs to the host server via PCIe endpoints,
GPUs are physically decoupled from host servers
and put into separate GPU boxes.
Correspondingly, each GPU box contains several GPUs.
To enable the remote access to GPUs from the host server, a proxy, named \proxyname,
is added on both the server and the GPU side.
It acts as medium for the interactions between the host server and GPUs.
The \proxyname is responsible for \textit{the conversion between PCIe TLP and network packets}.
On the one hand, \proxyname provides PCIe interfaces to get connected with the host server or GPUs.
On the other hand, {\proxyname}s are interconnected via a network fabric protocol.
Specifically, when receiving PCIe TLP packets, \proxyname encapsulates them into network packets and sends out these packets.
Correspondingly, on the other side, \proxyname extracts the original TLP packets from the received network packets.
Therefore, \textit{the conversion between PCIe TLP and network packets} is transparent to both the host server and GPUs.
From the perspective of the host server, GPUs are still attached locally via the PCIe link.
The combination of {\proxyname}s at both sides works as a PCIe switch (Figure \ref{fig:logic_arch}).
Moreover, \proxyname provides configuration interfaces to enable the dynamic distribution of GPUs for the cloud orchestration system.


The key observation that motivates the adoption of the network fabric is that nowadays network cables can provide the same (or even faster) data transmission speed when compared with PCIe links. 
For example, a PCIe Gen 3 interface offers about 7.87Gbps per lane at the physical layer~\cite{pcie-bench}, and the PCIe interface provided by a GPU has no more than 16 lanes, which can be supported easily with two 100GbE ethernet cables.
What's more, the adoption of the network fabric can perfectly solve the \textit{scope} and \textit{capacity} problems mentioned above.

\noindent
\textbf{Supporting Datacenter-Scale GPU Disaggregation.}
In \sysname, servers and GPUs are connected to the network fabric through network cables, whose maximum length can be up to 100 metres~\cite{Ethernet_cable}.
Hence, the scope of \textit{GPU disaggregation} is no longer limited to the rack scale.
GPU nodes can be allocated to host servers from anywhere in the datacenter, enabling fine-grained and efficient GPU provisioning~\cite{Resource_disaggregation}.

\noindent
\textbf{Supporting Large Capacity GPU Pools.}
Driven by the developments of network technologies, nowadays, the number of ports~\cite{Huawei, cisco} in a datacenter-level network is enough for \sysname to hold more GPU nodes. 
Specifically, compared with the PCIe fabric, data transfer in the network fabric is more reliable in its error check (e.g., Receive Errors, Send Errors, and Collisions) and packet retransmission.
So the network fabric is more tolerant to transmission error and expansible, where we can simply cascade the switches to build a larger network.
With its help, a GPU pool in \sysname can accommodate hundreds of nodes.


It is to be noted that, in \sysname, GPUs are \textit{dedicated to the host server during use}.
The binding relationships between GPUs and the host server won't be changed unless users free them.



\subsection{GPU Box}
\label{sec:GPU_box}
With respect to GPU Boxes, there are two kinds of design schemes.
The first one learns from NVIDIA DGX Systems~\cite{DGX}, which specialize in GPGPU acceleration.
Under this scheme, GPUs and the motherboard containing \textit{NVSWITCH}es are welded together, leaving PCI interfaces for \proxyname{s}.
Thus \textit{NVLINK}s can be utilized in \sysname to support more efficient GPU communications.
The second one is a simple PCIe switch that we have mentioned above.
All kinds of GPUs that connected via PCI interfaces can be deployed and provisioned in this design.
The former design is suitable for users requesting multiple GPUs and high compute capability, where users should be allocated with nodes connected by \textit{NVLINK}s.
While the latter one is appropriate for single GPU request scenarios.

\begin{table}[t]
    \centering
		\begin{minipage}{.45\linewidth}
		\centering
		\caption{Content of Mapping Tables in \proxyname (on the host server side).}
		\label{tab:record_GPU_host_side}
		\footnotesize
		\begin{tabular}[b]{c}
			\hline
			Table Entry ID \\
			\hline
			Used (whether the bus has been used) \\
			\hline
			Bus ID (allocated by OS during enumeration) \\
			\hline
			Device ID \\
			\hline
			Memory Base (allocated by OS during enumeration) \\
			\hline
			Memory Limit (allocated by OS during enumeration) \\
			\hline
			GPU Box ID (allocated by \managername) \\
			\hline
			Slot ID (allocated by \managername) \\
			\hline
			Path ID (allocated by \managername) \\
			\hline
		\end{tabular}
    \end{minipage}
	\qquad
	\begin{minipage}{.45\linewidth}
		\centering
		\caption{Content of Mapping Tables in \proxyname (on the GPU Box side).}
		\label{tab:record_host_GPU_side}
		  \footnotesize
		\begin{tabular}[b]{c}
			\hline
			Table Entry ID \\
			\hline
			Valid (whether the GPU is in place) \\
			\hline
			Used (whether the GPU has been used) \\
			\hline
			Slot ID \\
			\hline
			Host Node ID (allocated by \managername) \\
			\hline
			Path ID (allocated by \managername) \\
			\hline
		\end{tabular}
		\end{minipage}
  \end{table}

\subsection{\proxyname}
\label{sec:proxy}
In \sysname, \proxyname acts as a \textit{PCIe virtual switch} for the interactions between the host server and GPUs.
From the host server and GPU side, {\proxyname}s are inserted into the PCIe slot of the host server and GPU Box.
And {\proxyname}s are interconnected via a network fabric protocol.
During the boot stage of host server, BIOS program only enumerates the \textit{PCIe virtual switch} displayed by \proxyname.
So it would reserve enough memory space for \textit{disaggregated GPUs}, which would be hotplugged into the host server later.

\subsubsection{Mapping Tables in \proxyname}
As can be seen from Table~\ref{tab:record_GPU_host_side} and Table~\ref{tab:record_host_GPU_side}, there are two types of mapping tables stored in \proxyname on the host server and GPU Box side, respectively.
These tables record the mapping relationships between host servers and GPUs.
Here, we introduce \managername, which is responsible for GPU provisioning and reclaim in the fabric.
During the initialization stage of the network fabric, the node IDs of {\proxyname}s on each host server and GPU box are allocated by \managername automatically.

\noindent
\textbf{Mapping Tables On the Host Server Side.}
From the perspective of the host server, the \textit{disaggregated GPU} is a PCI device on the PCIe virtual switch displayed by the \proxyname.
So the mapping table must record the relationship between the PCI device on the host server (Bus ID, Device ID, Memory Base, Memory Limit) and the GPU (GPU box ID, slot ID).
When one host server or VM in the fabric send GPU requests to \managername, it would search for free GPUs and allocate them.
On the host server side, it would search free PCI bus in the mapping table, write the GPU Box, Slot, and path ID, then sets the used bit.
After detecting new PCI device, OS of the host server would enumerate the PCI bus tree again, allocating memory space and I/O resource for GPUs.
\proxyname would rewrite the value of memory base and limit in the mapping table based on the TLP configuration packets.

\noindent
\textbf{Mapping Tables On the GPU Box Side.}
Since each GPU is inserted into one slot in a GPU Box, the mapping table on the GPU Box side mainly records the slot ID of each GPU and its host server.
During the allocation of GPUs, \managername would search a free GPU in the fabric, write the host node, and path ID, then sets the used bit.

\subsubsection{Formats of Network Packets}
Since the \proxyname is responsible for \textit{the conversion between PCIe TLP and network packets},
we describe the basic formats of network packets to show the conversion process.
Generally, \proxyname would split header and data packets of original PCIe packets, and encapsulate them into network header and data packets, correspondingly.

\begin{figure*}[t]
	\centering
	
	\subfloat[Original PCIe Non-posted TLP Flow \label{fig:pcie_direct}]{
		\includegraphics[width=0.48\textwidth]{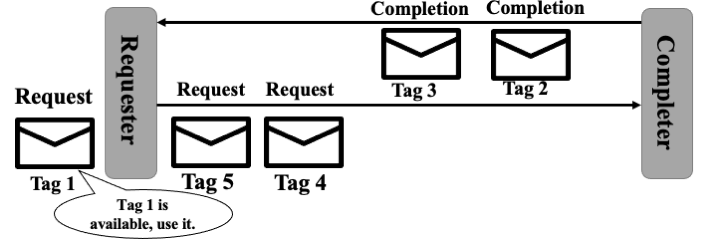}
	}
	\subfloat[PCIe Non-posted TLP Flow with \sysname \label{fig:pcie_pool}]{
		\includegraphics[width=0.48\textwidth]{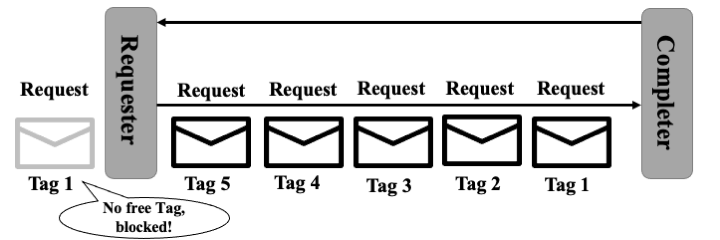}
	}

	\caption{Comparison of PCIe Non-posted Transaction TLP Flows. Assuming that the Requester maintains 5 tags,
			(a) shows that 5 tags are enough to keep the whole TLP flow working in pipeline,
			(b) shows that new transaction gets blocked due to the lack of free tags.
			This makes round-trip latency fail to be hidden and decreases the data transfer throughput.
			}
\end{figure*}

\noindent
\textbf{Formats of the Header Packets.}
The content of the network header packets can be split into network route information, PCIe TLP header, Cyclic Redundancy Check (CRC) code of TLP, CRC code of network packets.
In addition, the network route information should include source information(source node and slot ID~\footnote{For host server, this value should be 0.}), destination information(destination node and slot ID), and path ID.

\noindent
\textbf{Formats of the Data packets.}
For network data packets, they mainly contain original TLP packets and CRC code of network packets.

\subsubsection{Routing for Different Types of TLPs}
On the host server side, \proxyname should consider the routing for different types of PCIe TLPs.
For address-based routing transactions(e.g., Memory and I/O Transactions), \proxyname can retrieve the destination GPU box ID and slot ID by comparing the address in the TLP with (Memory Base, Memory Limit) in the mapping table.
For ID-based routing transactions(e.g., Configuration Transactions), \proxyname can get the target tuple via (Bus ID, Device ID).
For implicit routing transactions, \proxyname should firstly identify the message routing value in TLPs and categorize them into address-based, ID-based routing, or other local transactions.
On the GPU side, as a GPU is only dedicated to one host server during use, \proxyname just needs to encapsulate PCIe packets into network packets directly.

\subsection{Performance Model}
\label{sec:performance_model}

In this section, we start by explaining how network latency incurs performance overhead.
To measure this impact and guide the implementation of \sysname, we build up a performance model, which is based on AI workloads.
It is to be noted that, here the performance denotes \textit{GPU compute capability} in \sysname compared with the native one, which is different from the model prediction accuracy in AI scenarios.

Despite the comparable data transmission rate with a network cable,
\sysname still introduces non-negligible latency in the PCIe TLP transfer between the host server and GPUs.
Furthermore, the introduced latency may affect the throughput of the data transfer between them for some PCIe TLP transactions.

PCIe TLP transactions are categorized into non-posted (e.g., Memory reads) and posted transactions (e.g., Memory Writes), depending on whether the Request (i.e., the issuer of the transaction) expects completion TLPs in a transaction.
In non-posted transactions, the Requester won't finish the request until they receive one or more completion TLPs from Completer (i.e., the receiver of the transaction request).
Therefore, the Requester needs to trace the status of all in-flight non-posted transactions.
To this end, the Requester assigns each non-posted transaction with an identification (named \textit{tag})
and is responsible for ensuring the uniqueness of \textit{tag} for each in-flight non-posted transaction to distinguish them from each other.
When there are enormous data transfers, the number of tags supported, which is device-specific, is generally enough. 
And new non-posted transactions won't be blocked (Figure~\ref{fig:pcie_direct}).
However, as shown in Figure~\ref{fig:pcie_pool}, the latency introduced by \sysname makes the tags used up quickly.
New transactions are postponed until a free tag is available.
As a result, the throughput of data transfer based on non-posted TLP transactions (e.g., DMA read issued by GPUs, which is implemented via PCIe Memory Read) would decrease significantly.

As we have discussed, \sysname brings the longer latency for PCIe TLP transfers \ding{182} and lower throughput for non-posted PCIe TLP transactions \ding{183}.
The next question arising is: \textit{What is the performance impact of \sysname for GPU workloads?}
To answer this question, we build up a performance model via API hooking techniques to
simulate the introduced latency by \sysname (\ding{182}\ding{183}). 
Moreover, our performance model targets AI workloads, since they cover the most of GPU application scenarios in the cloud.

\subsubsection{\textit{Design of Performance Model}}
\label{sec:design_performance_model}

We analyze the main functions of AI workloads that happen during interactions between the host server and GPUs, and categorize them into three aspects:
\begin{itemize}[nosep,leftmargin=1em,labelwidth=*,align=left]

	\item \textbf{Memcpy(HtoD).}
	It means the data transfer from host memory to GPU memory.
	It is executed via issuing DMA read operations by GPUs,
	which would generate PCIe Memory Read transaction(s) correspondingly.
	Memcpy(HtoD) is mainly used to transfer input data, so that GPU kernels can consume.

	\item \textbf{Memcpy(DtoH).}
	It means the data transfer from GPU memory to host memory.
	It is executed via issuing DMA write operations by GPUs.
	Correspondingly, PCIe Memory Write transaction(s) are generated.
	Memcpy(DtoH) is mainly used to transfer the GPU kernel's result back.

	\item \textbf{Kernel Execution.}
    GPU kernels are functions executed in GPUs.
	When a GPU kernel finishes execution, it needs to access host memory, 
	which would generate PCIe Memory Read transaction(s) correspondingly.
\end{itemize}

We then summarize how \ding{182} and \ding{183} affect the above aspects as follows.
For convenience, we introduce some denotations.
$RTT_{\sysname}$ represents the round-trip latency of PCIe memory read transaction with \sysname,
while $RdTP_{\sysname}$ represents the lower throughput of PCIe memory read transaction with \sysname.
Similar definitions also apply to $RTT_{ori}$ and $RdTP_{ori}$ for non-disaggregated architecture.
Moreover, $RTT_{delta}$ denotes $RTT_{\sysname}$ minus $RTT_{ori}$.

\begin{itemize}[nosep,leftmargin=1em,labelwidth=*,align=left]
    
	\item \textbf{Memcpy(HtoD).}
	There are two scenarios since it is based on non-posted transactions.  
	First, if the data being transferred does not consume all tags, there will be just $RTT_{delta}$ overhead introduced \ding{182}.
	Second, if there is a large mount of data being transferred, the tags would quickly be depleted, and new transactions would get blocked. 
	As a result, the data would be transferred at a rate of $RdTP_{\sysname}$ \ding{183}.

	\item \textbf{Memcpy(DtoH).}
	Since it is realized on posted transactions, only \ding{182} would cause effect and its bandwidth is not affected by \ding{183}.
	Moreover, considering no completion TLP is needed, the introduced latency overhead can be estimated to be about 0.5 $RTT_{delta}$.

	\item \textbf{Kernel Execution.}
	There would be only \ding{182} that cause effect and introduces $RTT_{delta}$ overhead.
\end{itemize}

In fact, the quantitative relationship between $RdTP_{\sysname}$ and $RTT_{\sysname}$ can be estimated using the following formula:
\begin{equation}
	\label{eq:1}
	RdTP_{\sysname} \; = \; \#tags \: * \: MRS \: / \: RTT_{\sysname}
\end{equation}
where \textit{\#tags} denotes the number of tags supported by a GPU. 
\textit{MRS} represents \textit{Max\_Read\_Request\_Size}, which determines the maximal data size that the Requester can request in a single PCIe read request.


\subsubsection{\textit{Implementation of Performance Model}}
\label{sec:implementation_performance_model}

We implement our performance model by hooking CUDA Driver API to inject the effect of \ding{182} and \ding{183}, respectively.

\begin{itemize}[nosep,leftmargin=1em,labelwidth=*,align=left]

	\item \textbf{Memcpy(HtoD).} 
	We hook all related CUDA Driver APIs for Memcpy(HtoD), including synchronous (e.g., cuMemcpyHtoD) and asynchronous (e.g., cuMemcpyHtoDAsync).
	For small data, we insert one more Memcpy(HtoD) operation to copy a fix-sized buffer,
	where the time spent would be $RTT_{delta}$.
	For large data transfers, we would insert additional Memcpy(HtoD) operations to increase the time it takes for data copy to \textit{$RdTP_{ori\xspace}$ / $RdTP_{\sysname}$} times.
	In this way, we can simulate the time effect of \ding{183}.

	\item \textbf{Memcpy(DtoH).}
	All related CUDA Driver APIs for (synchronous or asynchronous) Memcpy(DtoH) are hooked, such as cuMemcpyDtoH and cuMemcpyDtoHAsync.
	In the hooker, we insert one more Memcpy(DtoH) operation, 
	which would copy a fix-sized buffer and time spent would be half of $RTT_{delta}$.

	\item \textbf{Kernel Execution.}
	We hook the related CUDA Driver APIs for kernel launch (e.g., cuLaunchKernel).
	We launch another dummy GPU kernel before the original GPU kernel.
	The execution time of the dummy GPU kernel is equal to $RTT_{delta}$.
	Moreover, since the execution of memset shares similar behavior with the kernel execution,
	we hook memset related CUDA Driver APIs and also insert the dummy GPU kernel.
\end{itemize}

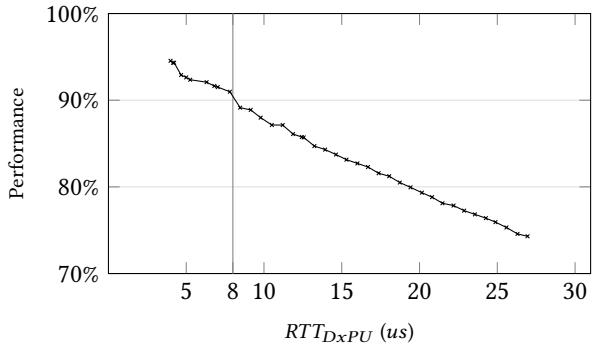
\begin{figure}[t]
    \begin{tikzpicture}
        \begin{axis}[
                xmin=0,
                xmax=31,
                xtick={5, 10, 15, 20, 25, 30},
                xticklabels={5, 10, 15, 20, 25, 30},
                xlabel = {$RTT_{\sysname}$ ($us$)},
                ymin=0.7,
                ymax=1.0,
                ytick={0.7,0.8,0.9,1},
                yticklabels={70\%,80\%,90\%,100\%},
                ylabel={Performance},
                ylabel style={font=\footnotesize},
                xlabel style={font=\footnotesize},
                extra x ticks={8},
                extra x tick style={grid=major,major grid style={gray}},
                ymajorgrids=true,
                minor grid style={gray!25},
                major grid style={gray!25},
                legend style={nodes={scale=0.6, transform shape}},
                legend pos=south east,
                legend cell align = {left},
                height = 5cm,
                width = 8cm,
            ]
            \addplot[mark=x, mark size=1pt]
                table[x=latency,y=perf,col sep=comma]{./data/sim_data.csv};
        \end{axis}
    \end{tikzpicture}
    \caption{Results of Performance Model for ResNet-50.
	To ensure that the performance is above 90\%, $RTT_{\sysname}$ needs to be within 8us.
	}
    \label{fig:perf_res}
\end{figure}

\subsubsection{\textit{Results of Performance Model}}
\label{sec:result_performance_model}

We run our performance model on a server with a single GPU (NVIDIA Tesla V100 SXM2 16GB).
ResNet-50 (AI framework: TensorFlow 1.15) is chosen as our target AI workload, considering its representative and popularity.
In addition, our experiment shows that the number of effective in-flight \textit{\#tags} in a GPU is about 140, while \textit{MRS} is 128.

Figure~\ref{fig:perf_res} demonstrates the relationship between $RTT_{\sysname}$ and the AI workload performance.
We can see that the performance of AI workload decreases with the increase of $RTT_{\sysname}$.
Specifically, when $RTT_{\sysname}$ is about 19us, the performance would decrease to 80\%.
Meanwhile, to ensure that the performance is above 90\%, $RTT_{\sysname}$ needs to be within 8us.

\subsubsection{\textit{Limitations of Performance Model}}
\label{sec:limitation_performance_model}

Since our performance model is based on the round-trip latency between the host server and GPUs, it cannot be directly applied to communications between GPUs.
Additionally, in our cases, there are three types of scenarios for GPU-to-GPU interactions, which involve NVLinks, PCIe Links, and \proxyname{s}.
Thus, it is difficult to cover all of them and make accurate modeling in multi-GPU scenarios.
Moreover, model setup via API hooking brings fluctuations and can not cover all interactions functions between the host server and GPUs.
We consider multi-GPU support and more accurate performance modeling (e.g., via cycle-level GPU simulator, or GPGPUsim) as our future work.

\subsection{Implementation of \sysname}
\label{sec:implementation}
According to results of performance model, 
we notice that the training performance of typical ResNet-50 model can still remain above 90\% with about 8us $RTT_{\sysname}$.
This motivates us to further prototype the design of \sysname.
There are two kinds of straightforward options for us to build the implementation system of \sysname.
The first option is leveraging off-the-shelf PCIe-over-TCP commercial products.
However, these products target low-speed devices and thus are not suitable due to the performance requirement mentioned above.
The second option is building by ourselves.
Obviously, building such a product that achieves the product-ready status brings huge development burden and hinders our prototype process.
Instead, we present the above design to our third-party vendors and seek for a satisfying product to act as \proxyname.

As a result, we build two customization-level implementation systems of \sysname,
whose $RTT_{\sysname}s$ are 6.8us and 4.9us, respectively.
According to our performance model, such $RTT_{\sysname}s$ keep the performance of AI workloads in GPUs above 90\%.
Our systems implement a proprietary protocol for \textit{the conversion between PCIe TLP and network packets},
which is neither PCIe over TCP nor RoCE (RDMA over Converged Network).
From the host server and GPU side, {\proxyname}s are inserted into the PCIe slot of the host server and GPU Box.
And {\proxyname}s are interconnected via the protocol.
The PCIe interface for a GPU is PCIe Gen 3x16, while the two 100GbE network cables are used for each \proxyname to get interconnected.
Note that the interconnection between {\proxyname}s is not limited to the network cable in design.
For example, another different implementations could be building the network fabric based on Infiniband RDMA technology.
Furthermore, The implementation systems demonstrate the practicality of  \sysname and verify the accuracy of our performance model.

\subsubsection{Verification of our performance model}
We evaluate the throughput of our implementation systems with the \textit{bandwidthTest} tool provided as part of CUDA code samples,
which turns out to be 2.7GB/s (for 6.8us) and 3.9GB/s (for 4.9us).
According to Equation \ref{eq:1}, $RdTP_{\sysname}s$ are 2.64GB/s and 3.66GB/s, respectively.
The accuracy of Equation \ref{eq:1} is verified.

\begin{table}[tbp]
    \centering
    \caption{Performance Model Result Validation in ResNet-50 Training. The estimated results are close to performance of the implementation system, verifying the correctness of the performance model.
    }
    \label{tab:val}
	\small
    \begin{tabular}[b]{ccc}
        \hline
        $\bm{RTT_{\bm{DxPU}}}$& \textbf{Results of Preformance Model} & \textbf{Performance of Implementation System}\\
        \hline
        4.9us & 92.56\% & 91.50\% \\
        \cline{1-3}
        6.8us & 91.40\% & 89.56\% \\
        \hline
    \end{tabular}
\end{table}

\smallskip
\noindent
\textbf{Accuracy of Performance Model.}
As shown in Table \ref{tab:val},
the training performance of ResNet-50 under our implementation systems are about 89.56\% (for 6.8us), and 91.5\% (for 4.9us).
Correspondingly, the estimated results of our performance model are 91.4\% (for 6.8us) and 92.56\% (for 4.9us).
The accuracy of our performance model is verified.
Moreover, we can see our estimated results are a little higher than the real scenario.
We think it is because our performance model only targets the main interaction functions,
while the other interaction functions, such as event synchronization, are ignored.

\section{Evaluation}
  
\label{sec:evaluation}


As we have mentioned above, compared with solutions at the software layer, \sysname is excellent in \textit{full-scenario support}, \textit{software transparency}, and \textit{good performance}.
Compared with existing PCIe fabric based solutions, \sysname supports both \textit{datacenter-scale GPU disaggregation} and \textit{large capacity GPU pools}.
Hence, in the evaluation, we mainly analyze the performance overhead incurred by \sysname and seek to answer the following three research questions (RQs).
\begin{itemize}[nosep,leftmargin=2.5em,labelwidth=*,align=left]
	\item[\textbf{RQ1}] \textbf{What causes the performance overhead in \sysname?}
 
	\item[\textbf{RQ2}] \textbf{How different parameter configurations affect the performance overhead?}
 
    \item[\textbf{RQ3}] \textbf{What causes the extra performance overhead in the multi-GPU environment?}
\end{itemize}
To begin with, we analyze the components of $RTT_{\sysname}$ and show its impact on the bandwidth between the host server and GPUs (Section~\ref{sec:evaluation_comp}).
Then, we conduct detailed experiments to analyze the performance overhead (Section~\ref{sec:evaluation_perf}) with typical user cases in the cloud, and present its limitations (Section~\ref{sec:evaluation_limit}).
The results show that the performance overhead is no more than 10\% in most cases and proves feasibility of \sysname.
Note that although \textit{the conversion between PCIe TLPs and network packets} is applicable to other PCIe devices,
currently, the performance model and implementation systems are designed for GPU workloads.
And disaggregation of other PCIe devices in this way is considered as our future work.

\subsection{Experiment Setup}
We set up a \textit{disaggregated GPU} pool consisting of 32 NVIDIA GPUs (16 Tesla V100 GPUs in the first type GPU boxes and 16 RTX 6000 GPUs in the second type GPU boxes), coupled with 4 host servers.
In addition, the $RTT_{\sysname}s$ is 6.8us.
And the detailed configuration can be seen from Table~\ref{tab:server_config}.

\begin{table}[tbp]
    \begin{minipage}{.45\linewidth}
        \caption{Experiment Configuration. 
        The native environment is the server-centric architecture.
        GPUs in \sysname are selected from the resource pool.
        Other configurations are the same.
        }
        \label{tab:server_config}
        \centering
        \footnotesize
        \begin{tabular}[b]{ccc}
            \hline
            CPU&\multicolumn{2}{c}{Intel Xeon Platinum 8163} \\
            \hline
            \multirow{5}*{GPU}&\multicolumn{2}{c}{Basic and AI Workloads:} \\
            ~&\multicolumn{2}{c}{NVIDIA Tesla V100 SXM2 16GB}\\
            ~&\multicolumn{2}{c}{(supporting \textit{NVLINK}s)}\\
            \cline{2-3}
            ~&\multicolumn{2}{c}{Graphics Rendering Workloads:}\\
            ~&\multicolumn{2}{c}{NVIDIA GeForce RTX 6000 24GB}\\
            \hline
            Memory&\multicolumn{2}{c}{768GB} \\
            \hline
            Storage&\multicolumn{2}{c}{1200GB}\\
            \hline
            \multirow{2}*{Network}&\multirow{2}*{\sysname}&200Gbit/s Fabric Network\\ 
            \cline{3-3}
            ~&~&Single-hop Switches\\
            \hline
        \end{tabular}
    \end{minipage}
    \hspace{0.1in}
    \begin{minipage}{.45\linewidth}
    \centering
    \begin{threeparttable}
    \caption{Source of $RTT_{\sysname}$. Time latency incurred by \sysname is around 5.6us, increasing communication latency between the host server and GPUs.}
    \label{tab:components}
	\footnotesize
    \begin{tabular}[b]{ccc}
        \hline
        ~&Time Latency&Proportion \\
        \hline
        \textbf{Original Time Latency} & 1.2us&17.7\% \\
        \hline
        \textbf{Network Transmission} & 1.9us&27.9\% \\
        \hline
        \textbf{Packet Conversion}&3.7us & 54.4\% \\
        \hline
    \end{tabular}
    \end{threeparttable}

    \vspace{0.2in}

    \begin{threeparttable}
        \caption{Bandwidth between GPUs and Host servers based on CUDA bandwidthTest. Restricted by non-posted transactions, the bandwidth between the host server and GPU drops rapidly.}
        \label{tab:bandwidthh2d}
        \footnotesize
        \begin{tabular}[b]{cccc}
            \hline
            ~&\sysname&Native&Proportion \\
            \hline
            \textbf{Host Server to GPU} & 2.7GB/s&11.2GB/s&24.1\% \\
            \hline
            \textbf{GPU to Host Server} & 11.6GB/s&12.5GB/s&92.8\% \\
            \hline
        \end{tabular}
        \end{threeparttable}
    \end{minipage}
\end{table}

\subsection{Components of $RTT_{\sysname}$}
\label{sec:evaluation_comp}
To measure the components of $RTT_{\sysname}$, we request a single Tesla V100 GPU from the pool and perform repeated read-write operations in the GPU memory space.
During these operations, we calculate the time length of different parts in $RTT_{\sysname}$.
Generally, $RTT_{\sysname}$ can be split into three parts: original time latency, network transmission, and packet conversion.
As can be seen from Table~\ref{tab:components}, they make up 17.7\%, 27.9\%, and 54.4\% of $RTT_{\sysname}$, respectively.
The original time latency pertains to the duration taken to transmit PCIe packets without \sysname, where GPUs are PCIe devices on the host server.
The network transmission time denotes the time required to transmit network packets between {\proxyname}s.
The packet conversion time indicates the duration required to convert PCIe packets to network packets and any additional time spent due to the design of the network protocol (such as error control and redundancy check).

To demonstrate the impact of $RTT_{\sysname}$ on the bandwidth between the host server and GPUs, we record their values in Table~\ref{tab:bandwidthh2d}.
The bandwidth from the host server to GPUs experiences a rapid drop to only 24.1\% due to non-posted transactions, whereas the bandwidth from GPUs to the host server remains largely unaffected.

\subsection{Performance of \sysname}
\label{sec:evaluation_perf}

From the GPU pool, We request a single RTX 6000 GPU for basic workloads, 8 Tesla V100 GPUs (supporting \textit{NVLINK}s) for AI workloads, and a single RTX 6000 GPU for graphics rendering workloads.
We compare the evaluation results in \sysname with those in the native environment, whose architecture is server-centric.
And other configurations of \sysname and the native environment are the same.
It is to be noted that, here the performance denotes \textit{GPU compute capability} in \sysname compared with the native one, which is different from the model prediction accuracy in AI scenarios.

We aim at measuring the performance impact brought by \sysname on the typical user cases in the cloud.
To this end, we adopt the following standard benchmarks:
\begin{itemize}[nosep,leftmargin=1em,labelwidth=*,align=left]
	\item \textbf{Basic Workloads: NVIDIA CUDA Samples.}
    CUDA Samples~\cite{CUDA_sample} are NVIDIA official benchmarks that can be used to measure GPU performance.
    Here, we mainly use them to test \sysname for several basic workloads(e.g., add, copy, read, and write) and have a preliminary understanding of its performance.
    \smallskip
	\item \textbf{AI Workloads: ResNet-50 and BERT.}
    ResNet-50 (image classification), BERT (language modeling), SSD320 (object detection), and NCF (recommedation) are popular benchmarks in AI scenarios.
    We use tools from \textit{DeepLearningExamples}~\cite{DeepLearningExamples} and \textit{tf\_cnn\_benchmarks}~\cite{tfcnnbenchmarks}.
    \textit{TensorFlow 21.02-tf2-py3 NGC container} is the running environment~\cite{container}.
    \item \textbf{Graphics Rendering Workloads: valley, heaven, and glmark2.}
    \textit{valley}~\cite{valley}, \textit{heaven}~\cite{heaven}, and \textit{glmark2}~\cite{glmark2} are well-known benchmarks that simulate real-world graphics rendering applications,
\end{itemize}

\begin{table}[tbp]
    \centering
    \caption{Performance of \sysname in Basic Workloads with a Single GPU. 
    The peformance is better than 96\% in these common operations.
    }
    \label{tab:perf_basic}
    \footnotesize
    \begin{tabular}[b]{cccccccccccc}
        \hline
        \multicolumn{3}{c}{\textbf{General Matrix Multiplication}}&\multicolumn{3}{c}{\textbf{Fast Fourier Transform}}& \multicolumn{6}{c}{\textbf{Stream Operations}}\\
        \hline
        \textbf{FP16}&\textbf{FP32}&\textbf{FP64}&\textbf{FP16}&\textbf{FP32}&\textbf{FP64}&\textbf{Copy}&\textbf{Scale}&\textbf{Add}&\textbf{Triad}~\tnote{1}&\textbf{Read}&\textbf{Write} \\
        \hline
        97.2\%&99.5\%&99.1\%&96.3\%&97.5\%&98.3\%& 98.8\%&99.1\%&99.3\%&99.3\%&97.3\%&97.5\%\\
        \hline
    \end{tabular}
\end{table}

\subsubsection{\textit{Performance in Basic Workloads}}
According to NVIDIA official benchmark documents, we select commonly used operations and record the performance of \sysname in these basic operations.
As can be seen from Table~\ref{tab:perf_basic}, the test cases cover general matrix multiplication, fast fourier transform, and stream operations (i.e., copy, scale, add, triad, read, and write).
The performance overhead is no more than 4\% in all basic test cases.

\subsubsection{\textit{Performance in AI Workloads}}
To answer all above research questions in AI scenarios, we show and analyze the performance overhead of \sysname in AI workloads for single GPU and multi-GPU environments.

To begin with, we conduct experiments in ResNet-50 utilizing a single GPU with diverse parameter configurations, where the performance overheads are different. 
Table~\ref{tab:AI_performance} shows that the best performance can reach 99.6\%.
The default parameter values are: batch size = 64, local parameter device = GPU, xla = off, mode = training, dataset = synthetic.
It can also be discovered that setting smaller batch size or CPU as local parameter device increases the overhead gradually. In the following paragraphs, we firstly figure out the components of performance overhead in the single GPU environment to answer \textbf{RQ1}.

\begin{table*}[tbp]
    \centering
    \begin{threeparttable}
    \caption{Performance of \sysname in ResNet-50 with a Single GPU. 
    The performance is better than 90\% in most cases with different parameters.
    }
    \label{tab:AI_performance}
    \scriptsize
    \begin{tabular}[b]{cccccccccc}
        \hline
        \multirow{3}*{\tabincell{c}{\textbf{Batch} \\ \textbf{Size}}}&\multirow{3}*{\tabincell{c}{\textbf{Local} \\ \textbf{Parameter} \\ \textbf{Device}}}&\multicolumn{4}{c}{\textbf{Synthetic Dataset}}&\multicolumn{4}{c}{\textbf{ImageNet Dataset}} \\
        \cline{3-10}
        ~&~&\multicolumn{2}{c}{\textbf{Training Performance}}&\multicolumn{2}{c}{\textbf{Inference Performance}}&\multicolumn{2}{c}{\textbf{Training Performance}}&\multicolumn{2}{c}{\textbf{Inference Performance}}\\
        \cline{3-10}
        ~&~&\textbf{XLA OFF}\tnote{1}&\textbf{XLA ON}&\textbf{XLA OFF}&\textbf{XLA ON}&\textbf{XLA OFF}&\textbf{XLA ON}&\textbf{XLA OFF}&\textbf{XLA ON}\\
        \hline
        32&GPU&85.2\%&93.3\%&95.4\%&94.6\%&83.8\%&92.4\%&94.1\%&93.2\%\\
        \hline
        64&GPU&92.7\%&97.5\%&98.6\%&98.1\%&89.4\%&96.8\%&97.2\%&97.1\%\\
        \hline
        \multirow{2}*{128}&GPU&95.5\%&97.9\%&99.5\%&99.6\%&94.3\%&97.3\%&97.7\%&97.6\%\\
        \cline{2-10}
        ~&\cellcolor{gray!40}CPU\tnote{2}&\cellcolor{gray!40}90.9\%&\cellcolor{gray!40}88.1\%&\cellcolor{gray!40}83.4\%&\cellcolor{gray!40}74.0\%&\cellcolor{gray!40}90.5\%&\cellcolor{gray!40}88.0\%&\cellcolor{gray!40}84.1\%&\cellcolor{gray!40}73.5\%\\
        \hline
    \end{tabular}
    \begin{tablenotes}
        \item[1] Accelerated linear algebra (xla) can be utilized for operation fusion.
        \item[2] In \sysname, setting CPU as local parameter service increases network communication overhead greatly.
        Thus, this parameter configuration is not recommended, and related statistics are not covered in the summary of performance.
        So we mark them gray.
    \end{tablenotes}
    \end{threeparttable}
\end{table*}

\noindent
\textbf{Single GPU Performance Analysis.}
Considering that {GPU disaggregation} in \sysname is implemented at the PCIe level, it is transparent to OS and applications.
Therefore, the differences between \sysname and the native environment are the \textit{bandwidth} and \textit{command latency}, which are generated by \textit{network latency} between the host server and GPUs (mentioned in Section~\ref{sec:performance_model}).
We utilize NVIDIA Nsight System~\cite{NVIDIANsightSystem} to conduct a detailed investigation.
All detailed data is collected from experiments on the native GPU servers.

\begin{figure*}[tbp]
    \subfloat[CDF of Kernel Number Proportions with Duration. The batch size doesn't affect the proportion of \textit{short-duration kernels} (smaller than or equal to 10us) greatly.\label{fig:kernel_num}]{
    \begin{tikzpicture}
        \begin{axis}[
            ymax = 120,
            xmax = 1300,
            xtick={10, 200, 400, 600, 800, 1000},
            ytick={0, 20, 40, 60, 80, 100},
            xlabel = {Single kernel duration (us)},
            ylabel={Cumulative Percentage for \\ Number Proportions (\%)},
            ylabel style={{align=center}, font=\footnotesize},
            xlabel style={font=\footnotesize},
            extra x ticks=10,
            extra x tick style={grid=major,major grid style={gray}},
            extra y ticks=60,
            extra y tick style={grid=major,major grid style={gray}},
            ymajorgrids=true,
            minor grid style={gray!25},
            major grid style={gray!25},
            legend style={nodes={scale=0.6, transform shape},at={(1,0.33)},anchor=north east},
            legend cell align = {left},
            height = 4.5cm,
            width = 6cm,
            ]
            \addplot[color=black!50!blue]
                table[x=duration,y=distribution,col sep=comma]{./data/AI/resnet50/32_cumulative_kernel.csv};
            \addlegendentry{batch size 32}
            \addplot[color=darkgray]
                table[x=duration,y=distribution,col sep=comma]{./data/AI/resnet50/64_cumulative_kernel.csv};
            \addlegendentry{batch size 64}
            \addplot[color=lightgray]
                table[x=duration,y=distribution,col sep=comma]{./data/AI/resnet50/128_cumulative_kernel.csv};
            \addlegendentry{batch size 128}
        \end{axis}
    \end{tikzpicture}
    }
    \qquad
    \subfloat[CDF of Kernel Time Proportions with Duration. A larger batch size causes the average \textit{kernel duration} to increase and improves the performance.\label{fig:kernel_duration}]{
        \begin{tikzpicture}
            \begin{axis}[
                    xmax= 1300,
                    ymax = 120,
                    xtick={10, 200, 400, 600, 800, 1000},
                    ytick={0, 20, 40, 60, 80, 100},
                    xlabel = {Single kernel duration (us)},
                    ylabel = {Cumulative Percentage for \\ Time Proportions (\%)},
                    ylabel style={{align=center}, font=\footnotesize},
                    xlabel style={font=\footnotesize},
                    extra x ticks={10, 200, 800},
                    extra x tick style={grid=major,major grid style={gray}},
                    ymajorgrids=true,
                    minor grid style={gray!25},
                    major grid style={gray!25},
                    legend style={nodes={scale=0.6, transform shape},at={(1,0.33)},anchor=north east},
                    legend cell align = {left},
                    height = 4.5cm,
                    width = 6cm,
                ]
                \addplot[color=black!50!blue]
                    table[x=duration,y=distribution,col sep=comma]{./data/AI/resnet50/32_CDF_kernel_sum.csv};
                \addlegendentry{batch size 32}
                \addplot[color=darkgray]
                    table[x=duration,y=distribution,col sep=comma]{./data/AI/resnet50/64_CDF_kernel_sum.csv};
                \addlegendentry{batch size 64}
                \addplot[color=lightgray]
                    table[x=duration,y=distribution,col sep=comma]{./data/AI/resnet50/128_CDF_kernel_sum.csv};
                \addlegendentry{batch size 128}
            \end{axis}
        \end{tikzpicture}
    }
    \caption{CDF of Kernel Number and Time Proportions with Duration in ResNet.}
\end{figure*}
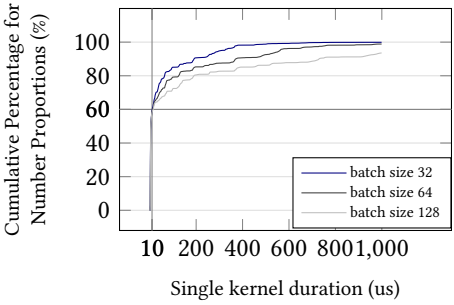
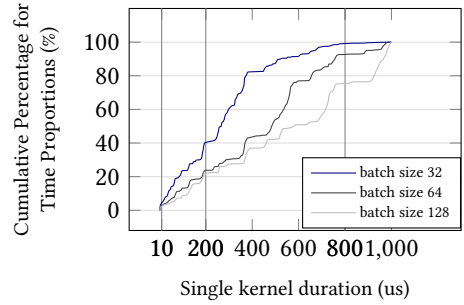

For \textit{bandwidth}, we collect the total transferred data between the host server and GPUs in each training step, which are only around 0.01MB and 40MB for the synthetic and ImageNet dataset.
Thus, the \textit{bandwidth} is not the critical cause of performance overhead in single GPU training or inference for \sysname. 
Therefore, we pay more attention to the effect of the \textit{command latency} in the following analysis.

Most of GPU workloads are made up of GPU kernel operations and memory load/store operations. 
With the default parameter configuration, GPU kernel execution makes up the most of GPU hardware time (around 99\%).
So we mainly analyze GPU kernel execution in these cases.
For simplicity, we categorize GPU kernels into \textit{short-duration kernels} (smaller than or equal to 10us) and \textit{long-duration kernels} (longer than 10us).
Then collected data shows that \textit{short-duration kernels} account for about 60\% of all kernels.
Obviously, the proportion of $RTT_{delta}$ is larger when the \textit{kernel duration} is smaller, especially for our cases where the percentage of \textit{short-duration kernels} is larger than 50\%.
Therefore, in our cases, the \textit{command latency} is the major factor that causes the performance decline in the single GPU environment.
What's more, it illustrates that users of \sysname should consider reducing the proportion of \textit{short-duration kernels} in their applications.

Moreover, it can be inferred from Table~\ref{tab:AI_performance} that different parameters influence the performance of \sysname greatly.
Therefore, we conduct a detailed analysis and explain the causes to answer \textbf{RQ2}.

\textit{Analysis of batch size:}
We make experiments with different batch sizes based on ResNet-50 synthetic dataset training.
Similar to the above findings, the proportions of memory operations are still no more than 1\% for each batch size, revealing that the batch size mainly affects \textit{kernel duration}.
From the above analysis we know that more than half of GPU kernels are \textit{short-duration kernels}.
If the percentage of \textit{short-duration kernels} increases, the percentage of $RTT_{delta}$ increases correspondingly, further worsening the performance.
Thus, in Figure~\ref{fig:kernel_num}, we plot the Cumulative Distribution Function (CDF) based on the number of the same duration kernels in all kernels.
However, the proportions of \textit{short-duration kernels} are 59.3\%, 58.9\%, 58.3\% when the batch sizes are 32, 64, 128.
Therefore, the batch size doesn't have significant effects on the proportion of \textit{short-duration kernels}.

Yet, further analysis shows that the \textit{long-duration kernels} are key factors influencing the performance.
We display the CDF based on the total time spent on the same duration kernels in all of time in Figure~\ref{fig:kernel_duration}.
It can be concluded that the \textit{long-duration kernels} influence the average value greatly.
For example, kernels ranging from 200-800us account for 58.9\%, 68.8\%, 53.6\% of the total \textit{kernel duration} when the batch sizes are 32, 64, 128, respectively.
In addition, the average \textit{kernel duration} are 56.0us, 102.3us, 193.0us correspondingly.
So in ResNet-50, a larger batch size causes the average \textit{kernel duration} to increase, further reducing the percentage of $RTT_{delta}$ and improving the performance.

\begin{table}[tbp]
    \begin{minipage}{.45\linewidth}
    \centering
    \caption{Proportions of Memory Operations in GPU Hardware Workloads with Different Local Parameter Devices(batch size = 128).
    The value increases rapidly when changing the parameter device to CPU.
    }
    \label{tab:AI_parameter_server}
    \footnotesize
    \begin{tabular}[b]{cccc}
        \hline
        \multirow{2}*{\textbf{Dataset}}&\multirow{2}*{\textbf{Mode}}&\multicolumn{2}{c}{\textbf{Local Parameter Device}}  \\
        \cline{3-4}
        ~&~&\tabincell{c}{\textbf{\qquad GPU}}&\tabincell{c}{\textbf{CPU}} \\
        \hline
        \multirow{2}*{Synthetic}&Training& \qquad 0.2\%&7.5\% \\
        \cline{2-4}
        ~&Inference& \qquad 0.2\%&15.6\% \\
        \hline
        \multirow{2}*{ImageNet}&Training& \qquad 2.8\%&9.9\% \\
        \cline{2-4}
        ~&Inference& \qquad 9.0\%&22.3\% \\
        \hline
    \end{tabular}
    \end{minipage}
    \hspace{0.1in}
    \begin{minipage}{.45\linewidth}
        \centering


        

        \begin{threeparttable}
            \caption{{\sysname}'s Performance in Object Detection and Recommedation Scenarios.
            }
            \label{tab:performance_object_detection_recommedation}
            \footnotesize
            \begin{tabular}[b]{ccccc}
                \hline
                \multirow{3}*{\textbf{SSD320 v1.2}} &\multicolumn{4}{c}{\textbf{Batch Size}} \\
                \cline{2-5}
                ~ & 8 & 16 & 32 & 64\\
                \cline{2-5}
                ~&81.53\%&84.62\%&84.47\%&83.64\%\\
                \hline
                \multirow{3}*{\textbf{NCF}} &\multicolumn{4}{c}{\textbf{Batch Size}} \\
                \cline{2-5}
                ~ & 16384 & 65536 & 262144 & 1048576 \\
                \cline{2-5}
                ~&96.78\%&98.04\%&97.55\%&98.25\% \\
                \hline
            \end{tabular}      
            \end{threeparttable} 
    \end{minipage}
\end{table}

\textit{Analysis of local parameter device:}
In terms of local parameter device, if the parameter device is set to CPU, more parameter-related operations are executed between the host server and GPUs.
Correspondingly, the proportion of $RTT_{delta}$ increases and the performance in Table~\ref{tab:AI_performance} decreases.
To demonstrate it in statistics, we collect proportions of memory operations with different local parameter devices in Table~\ref{tab:AI_parameter_server}.
It can be seen that the value increases rapidly when changing the parameter device to CPU, no matter which dataset we use.

\textit{Analysis of xla:}
Accelerated linear algebra (xla) is a domain-specific compiler which realizes the kernel fusion in the computational graph optimization.
Accordingly, the collected data shows that the average \textit{kernel duration} in the default situation increases from 102.3us to 131.0us when xla is on, thus improving the overall performance.

However, in Table~\ref{tab:AI_performance}, turning on xla worsens the performance when local parameter device is CPU or the mode is inference.
So we do a further analysis in these cases and record the changes.
For the former situation, we observe that the proportion of memory operations increases from 14.9\% to 16.7\%, and average duration of them decrease from 35.1us to 11.3us, 
all incurring more performance overhead.
For the latter one, we find that no matter in which case, turning xla on would increase the number and duration of memory operations.
In training cases, this increase can be ignored when compared with the total GPU kernels.
Yet, in inference cases, the number of kernels is far smaller than that in training.
And the proportion of memory operations increases from 0.2\% to 7.9\%, showing that it cannot be overlooked.

\textit{Analysis of mode and dataset:}
With regard to mode and dataset, we discover that the proportions of kernels and memory operations are the same in training and inference, generally.
On the other hand, the average \textit{kernel duration} increases in inference.
Statistically speaking, the average value increase by 48.2\%, 55.5\%, 55.1\%, compared with training cases, when the batch sizes are 32, 64, 128.
Therefore, the performance is better in inference.
As the ImageNet dataset is larger than the synthetic one in size, more transmission time between the host server and GPUs causes the proportion of $RTT_{delta}$ to increase.
Consequently, the performance overhead incurred by $RTT_{delta}$ increases a little bit for ImageNet dataset.

\begin{figure*}[tbp]
    \subfloat[CDF of Kernel Number Proportions with Duration. \label{fig:object_detection_kernel_num}]{
    \begin{tikzpicture}
        \begin{axis}[
            ymax = 120,
            xmax = 1300,
            xtick={10, 200, 400, 600, 800, 1000},
            ytick={0, 20, 40, 60, 80, 100},
            xlabel = {Single kernel duration (us)},
            ylabel={Cumulative Percentage for \\ Number Proportions (\%)},
            ylabel style={{align=center}, font=\footnotesize},
            xlabel style={font=\footnotesize},
            ymajorgrids=true,
            minor grid style={gray!25},
            major grid style={gray!25},
            legend style={nodes={scale=0.6, transform shape},at={(1,0.5)},anchor=north east},
            legend cell align = {left},
            height = 4.5cm,
            width = 6cm,
            ]
            \addplot[color=blue]
                table[x=duration,y=distribution,col sep=comma]{./data/AI/Detection/8_cumulative_kernel.csv};
            \addlegendentry{batch size 16}
            \addplot[color=black]
                table[x=duration,y=distribution,col sep=comma]{./data/AI/Detection/16_cumulative_kernel.csv};
            \addlegendentry{batch size 16}
            \addplot[color=red]
                table[x=duration,y=distribution,col sep=comma]{./data/AI/Detection/32_cumulative_kernel.csv};
            \addlegendentry{batch size 32}
            \addplot[color=green]
                table[x=duration,y=distribution,col sep=comma]{./data/AI/Detection/64_cumulative_kernel.csv};
            \addlegendentry{batch size 64}
        \end{axis}
    \end{tikzpicture}
    }
    \qquad
    \subfloat[CDF of Kernel Time Proportions with Duration.\label{fig:object_detection_kernel_duration}]{
        \begin{tikzpicture}
            \begin{axis}[
                    xmax= 1300,
                    ymax = 120,
                    xtick={10, 200, 400, 600, 800, 1000},
                    ytick={0, 20, 40, 60, 80, 100},
                    xlabel = {Single kernel duration (us)},
                    ylabel = {Cumulative Percentage for \\ Time Proportions (\%)},
                    ylabel style={{align=center}, font=\footnotesize},
                    xlabel style={font=\footnotesize},
                    ymajorgrids=true,
                    minor grid style={gray!25},
                    major grid style={gray!25},
                    legend style={nodes={scale=0.6, transform shape},at={(1,0.5)},anchor=north east},
                    legend cell align = {left},
                    height = 4.5cm,
                    width = 6cm,
                ]
                \addplot[color=blue]
                    table[x=duration,y=distribution,col sep=comma]{./data/AI/Detection/8_CDF_kernel_sum.csv};
                \addlegendentry{batch size 16}
                \addplot[color=black]
                    table[x=duration,y=distribution,col sep=comma]{./data/AI/Detection/16_CDF_kernel_sum.csv};
                \addlegendentry{batch size 16}
                \addplot[color=red]
                    table[x=duration,y=distribution,col sep=comma]{./data/AI/Detection/32_CDF_kernel_sum.csv};
                \addlegendentry{batch size 32}
                \addplot[color=green]
                    table[x=duration,y=distribution,col sep=comma]{./data/AI/Detection/64_CDF_kernel_sum.csv};
                \addlegendentry{batch size 64}
            \end{axis}
        \end{tikzpicture}
    }
    \caption{CDF of Kernel Number and Time Proportions with Duration in Object Detection Scenarios.}
    \label{fig:object_detection}
\end{figure*}
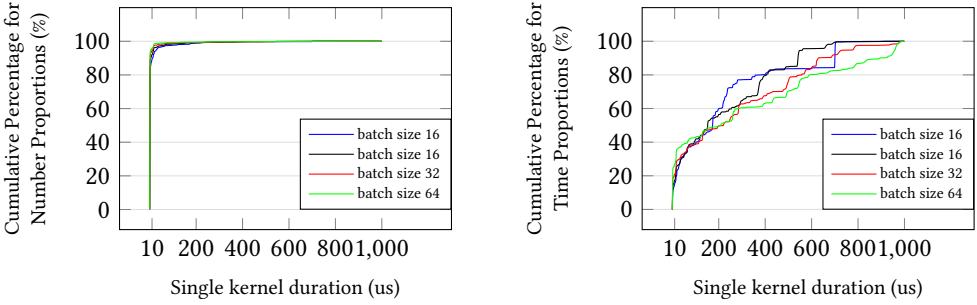

\textit{Analysis in Object Detection and Recommedation Scenarios:}
Table~\ref{tab:performance_object_detection_recommedation} demonstrates performance of \sysname in object detection and recommedation scenarios.
In test cases of NCF, the performance is above 96\% with different batch sizes.
However, when it comes to SSD320, the performance is around 83\%.
And we also plot the CDFs based on the number and time of kernels.
As can be seen from Figure~\ref{fig:object_detection}, the distributions of kernel durations with different batch sizes are similar.
In statistics, the average kernel durations for these cases are 10.7us, 8.2us, 7.9us, and 8.1us, respectively.
So the performance don't change a lot with different batch sizes.
Different from ResNet-50, the proportion of \textit{short-duration} kernels are more than 90\% with different batch sizes.
And this explains why its performance is only around 83\%.

\begin{table*}[tbp]
    \centering
    \caption{Bandwidth between the Host Server and GPU(s) in \sysname.
    The increase of \textit{bandwidth} doesn't follow the linear law when there are more than 4 GPUs, 
    showing that the communication bottleneck is triggered.
    }
    \label{tab:AI_bandwidth}
    \footnotesize
    \begin{tabular}[b]{ccccccc}
        \hline
        ~&\multicolumn{3}{c}{\textbf{Batch Size 64}}&\multicolumn{3}{c}{\textbf{Batch Size 128}} \\
        \hline
        ~&\textbf{Host to GPU(s)}&\textbf{GPU(s) to Host}&\textbf{Performance}&\textbf{Host to GPU(s)}&\textbf{GPU(s) to Host}&\textbf{Performance} \\
        \hline
        \textbf{1 GPU}&1.5GB/s&0.8GB/s&92.7\%&1.4GB/s&0.7GB/s&96.5\% \\
        \hline
        \textbf{2 GPUs}&2.6GB/s&1.3GB/s&90.3\%&2.6GB/s&1.3GB/s&94.7\% \\
        \hline
        \textbf{4 GPUs}&4.9GB/s&2.3GB/s&87.5\%&5.0GB/s&2.3GB/s&92.4\% \\
        \hline
        \textbf{8 GPUs}&8.4GB/s&3.6GB/s&82.4\%&8.3GB/s&3.6GB/s&90.7\% \\
        \hline
    \end{tabular}
\end{table*}

\captionsetup[subfloat]{justification=centering}
\begin{figure*}[tbp]
    \subfloat[Write Bandwidth \\(Unidirectional).]{
    \begin{tikzpicture}
        \begin{axis}[
            ymin=0,
            ymax=5,
            xbar,
            xmin=0,
            xmax=80,
            ytick={1,2,3,4},
            yticklabels={C1,C2,C3,C4},
            yticklabel style = {font=\footnotesize},
            xlabel={Bandwidth (GB/s)},
            ylabel style={{align=center}, font=\footnotesize},
            xlabel style={font=\footnotesize},
            ymajorgrids=true,
            minor grid style={gray!25},
            major grid style={gray!25},
            height = 4cm,
            width = 5cm,
            nodes near coords,
            nodes near coords align={horizontal},
            ]
            \addplot[draw=black,fill=white] coordinates {(9.4,1)(12.5,2)(24.2,3)(48.4,4)};
        \end{axis}
    \end{tikzpicture}
    }
    \subfloat[Write Bandwidth \\(Bidirectional).]{
        \begin{tikzpicture}
            \begin{axis}[
                ymin=0,
                ymax=5,
                xbar,
                xmin=0,
                xmax=140,
                ytick={1,2,3,4},
                yticklabels={C1,C2,C3,C4},
                yticklabel style = {font=\footnotesize},
                xlabel={Bandwidth (GB/s)},
                ylabel style={{align=center}, font=\footnotesize},
                xlabel style={font=\footnotesize},
                ymajorgrids=true,
                minor grid style={gray!25},
                major grid style={gray!25},
                height = 4cm,
                width = 5cm,
                nodes near coords,
                nodes near coords align={horizontal},
                ]
                \addplot[draw=black,fill=white] coordinates {(17.1,1)(22.6,2)(48.4,3)(96.5,4)};
            \end{axis}
        \end{tikzpicture}
    }
    \subfloat[Read Bandwidth.]{
        \begin{tikzpicture}
            \begin{axis}[
                ymin=0,
                ymax=5,
                xbar,
                xmin=0,
                xmax=80,
                ytick={1,2,3,4},
                yticklabels={C1,C2,C3,C4},
                yticklabel style = {font=\footnotesize},
                xlabel={Bandwidth (GB/s)},
                ylabel style={{align=center}, font=\footnotesize},
                xlabel style={font=\footnotesize},
                ymajorgrids=true,
                minor grid style={gray!25},
                major grid style={gray!25},
                height = 4cm,
                width = 5cm,
                nodes near coords,
                nodes near coords align={horizontal},
                ]
                \addplot[draw=black,fill=white] coordinates {(9.1,1)(12.5,2)(24.1,3)(47.8,4)};
            \end{axis}
        \end{tikzpicture}
    }
    \caption{Bandwidth between GPUs Based on CUDA p2pBandwidthLatencyTest (defaultly p2p enabled). 
    C1 means \proxyname. 
    C2 means the native environment. 
    C3 means GPUs are connected across one NVLINK.
    C4 means GPUs are connected across a bond settwo NVLINKs.
    GPUs in the native environment are communicated via a single PCIe bridge.
    In \sysname, GPUs under the same \proxyname are communicated through either PCIe bridges or NVLINKs, which relies on the design of GPU Boxes.}
    \label{fig:multi_GPU_bandwidth}
\end{figure*}
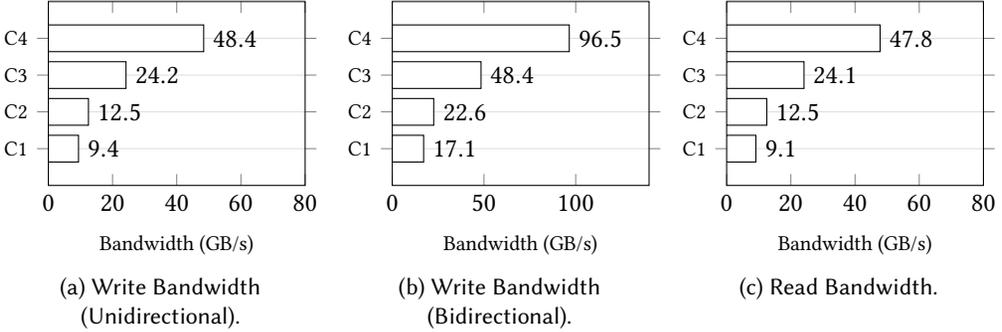

\begin{table}[!tbp]
    \begin{minipage}{.45\linewidth}
    \centering
    \caption{{\sysname}'s Performance in Graphics Rendering Scenarios.
    The performance of \sysname is acceptable (around 10\%) in real graphics rendering applications.}
    \label{tab:valley_heaven}
    \footnotesize
    \begin{tabular}[b]{ccc}
        \hline
        \textbf{Benchmark}&Valley&Heaven\\
        \hline
        \textbf{\sysname FPS}&2121.05&1870.29 \\
        \hline
        \textbf{Native FPS}&2177.67&2108.11\\
        \hline
        \textbf{Performance}&97.4\%&88.7\% \\
        \hline
    \end{tabular}
    \end{minipage}
    \hspace{0.1in}
    \begin{minipage}{.45\linewidth}
    \centering
    \caption{Average GPU Workload Duration and Performance in Glmark2 with Different Resolutions (object = ideas).
    The performance increases with a higher resolution.
    }
    \label{tab:ideas}
    \footnotesize
    \begin{tabular}[b]{ccc}
        \hline
        \textbf{Resolution}&\tabincell{c}{\textbf{Average GPU} \\ \textbf{Workload Duration}}&\textbf{Performance}\\
        \hline
        1920x1080&65.6us&87.9\%\\
        \hline
        3840x2160&122.8us&91.0\%\\
        \hline
        7680x4320&221.6us&93.0\%\\
        \hline
    \end{tabular}
    \end{minipage}
\end{table}

\noindent
\textbf{Multi-GPU Performance Analysis.}
Besides ResNet-50, we also conduct multi-GPU experiments in BERT (fine-tuning training for SQuAD v1.1).
We set the number of GPU(s) to 1, 4, and 8 and measure the performance overhead in multi-GPU scenarios to answer \textbf{RQ3}.
Since more GPUs incur more interactions (e.g., data, parameter, and synchronous operations) between the host server and GPUs,
the performance of \sysname declines when more GPUs are allocated.
In statistics, the performance of \sysname in Bert and ResNet are 94.6\%, 93.8\%, 93.4\% and 92.7\%, 87.5\%, 82.4\%, respectively.

\textit{Analysis of Host-server-to-GPU and GPU-to-Host-server bandwidth:}
In multi-GPU scenarios, besides \textit{command latency}, \textit{bandwidth} between the host server and GPUs may also affect the performance.
Since the amount of network packets, which \proxyname can process at the same time, is limited, a single \proxyname is not enough when more GPUs are allocated.
In order to measure this impact in statistics,
we change the number of GPUs and record the \textit{bandwidth} in Table~\ref{tab:AI_bandwidth}.
When the number is no more than 4, the \textit{bandwidth} increases at a linear speed.
However, when it switches from 4 to 8, the change of \textit{bandwidth} doesn't follow the law.
It reveals that the communication bottleneck is triggered.
Thus, in multi-GPU scenarios, users should take this effect into consideration and set up more \proxyname{s} according to the \textit{communication bandwidth}.
Additionally, if multiple host servers access GPUs in the same GPU Boxes, the bandwidth will increase quickly and communication bottleneck will be triggered, too.
In this case, users also need to deploy more \proxyname{s} according to the \textit{communication bandwidth}.

\textit{Analysis of GPU-to-GPU bandwidth:}
To demonstrate the bandwidth between GPUs across different {\proxyname}s, we test the corresponding PCIe read and write bandwidth.
As can be seen from Figure~\ref{fig:multi_GPU_bandwidth}, the bandwidth between GPUs across different {\proxyname}s is around 74\% of that across a single PCIe bridge.
Obviously, in multi-GPU cases, it is recommended that GPUs are allocated under the same \proxyname to avoid extra expense.
As we have mentioned that in our experiments, GPUs are connected via NVLINKs.
So bandwidth between GPUs aren't affected by \proxyname in these cases.

\subsubsection{\textit{Performance in Graphics Rendering Workloads}}
In this section, we test and analyze the performance of \sysname to answer \textbf{RQ1} and \textbf{RQ2} in graphics rendering scenarios.

According to Table~\ref{tab:valley_heaven}, performance overheads in \textit{valley} and \textit{heaven} are 2.6\% and 11.3\% respectively.
Therefore the performance of \sysname is acceptable in the real graphics rendering applications.

Moreover, we evaluate the performance of \sysname when it is assigned with specific tasks.
In glmark2, we select three targets (jellyfish, effect2d, ideas) and resolutions (1920x1080, 3840x2160, 7680x4320) for rendering.
And we discover that, the performance increases with a higher resolution.
To explain this phenomenon, we collect the duration of GPU workloads.
Typically the average GPU workload duration tends to be longer for OpenGL applications at higher resolutions.
Take ideas in Table~\ref{tab:ideas} as an example, the average values for these resolutions are 65.6us, 122.8us, 221.6us respectively.
So the performance of \sysname is better with a higher resolution.

To summarize, answers to research questions are as follows.
\begin{itemize}[nosep,leftmargin=2.5em,labelwidth=*,align=left]
	\item[\textbf{RQ1}] In our cases, \textit{command latency} is the major cause of the performance overhead incurred by \sysname.
	\item[\textbf{RQ2}] Effects of parameter configurations can be attributed to the changes in \textit{kernel duration} and proportions of memory operations. 
    \item[\textbf{RQ3}] \textit{Bandwidth} between the host server and GPUs affects performance overhead in multi-GPU scenarios greatly.
\end{itemize}

\subsection{Limitations of \sysname}
\label{sec:evaluation_limit}
As is mentioned above, in \sysname, the \textit{command latency} is longer, compared with the native environment.
Hence, \sysname is not appropriate for GPU workloads consisting of too many \textit{short-duration kernels} and memory operations.
Reducing the \textit{command latency} is also considered as our future work.

For users of \sysname, the proportions of \textit{short-duration kernels} and memory operations should be reduced in applications. 
And in multi-GPU scenarios, the number of \proxyname{s} should be considered based on the \textit{communication bandwidth}.

\section{Discussion and Future Work}
  
\label{sec:discussion}
\subsection{Software-Hardware Optimization Space}

\noindent
\textbf{Command Latency Overhead Mitigation.}
The effect caused by \textit{command latency} could be mitigated with two strategies:
interaction reduction and latency hiding.
For the former, it aims to reduce the interaction between the host server and GPUs.
For example, kernel fusion technique could be adopted to reduce the number of kernels executed.
For the latter, we could leverage concurrent and asynchronous techniques (such as streaming in CUDA) to perform multiple operations simultaneously and in pipeline.

\noindent
\textbf{Adoption of PCIe Gen 4.}
Currently, \sysname is built on PCIe Gen 3.
As PCIe Gen 4 supports the double throughput of PCIe Gen 3, the number of in-flight \#tags in GPUs would increase.
Correspondingly, the throughput of PCIe memory read transaction and memcpy(HtoD) would increase, too.
Meanwhile, communication efficiency and performance of \sysname in multi-GPU scenarios would also be better.
In the future, we will deploy \sysname on PCIe Gen 4.

\noindent
\textbf{PCIe Read Transaction Avoidance.}
As memcpy(HtoD) is built on PCIe Memory Read Transaction,
the throughput of memcpy(HtoD) in \sysname decreases dramatically (Ref. Equation~\ref{eq:1}).
To avoid this effect, an alternative memcpy(HtoD) implementation strategy can be adopted:
replacing PCIe Memory Read Transaction with PCIe Memory Write Transaction.
It can be achieved in two ways by leveraging hardware features of CPUs.
First, CPUs execute data move instructions to transfer data from host memory to GPU memory directly, 
which can be accelerated by leveraging the SIMD feature.
Second, we can leverage the DMA engine to finish the data transfer, which would issue PCIe Write Memory Transaction.
We implement a prototype for the first one
and the result shows that $RdTP_{\sysname}$ increases from 2.7 GB/s to 9.44 GB/s.

\subsection{GPU Distribution Scheme Design}
Although \textit{GPU disaggregation} is beneficial to cloud providers in resource utilization and elasticity,
the distribution scheme should also be well designed.
For example, GPUs in the same pool can be grouped and indexed in advance,
so user requirements can be satisfied quickly.
In addition, there should be spare GPUs in accordance with the equipment failure rate.
So broken GPUs can be replaced quickly to ensure that users have a good experience.

\subsection{End-to-end TLP Encryption}
As more and more users choose to store their private data in the cloud,
the importance of data privacy in the cloud is highlighted.
Thus, there have been many researches towards data privacy protection~\cite{succinct, SGXLock}.
In AI training and inference, the model and data are considered as user privacy, which is stored in TLPs during transmission.
In order to prevent the privacy leakage during network communication, end-to-end encryption can be applied in our future work.

\section{Related Work}
  
\subsection{Resource Disaggregation}
Resource disaggregation has attracted extensive attention from academia and industry due to its efficiency in resource management,
such as Intel Rack Scale Design~\cite{intelRack}, IBM Composable System~\cite{composableCS}, dReDBox~\cite{dredbox}, and so on.
However, although there are some studies trying to provide software support for them,
it is still challenging to adopt them in practice due to the compatibility or performance issue.

\subsection{GPU Virtualization}
With increasing demands for GPUs in the cloud, GPU virtualization is proposed to split compute capability of a single GPU server and allocate it in a more flexible way.
VOCL~\cite{VOCL}, rCUDA~\cite{rCUDA}, vCUDA~\cite{vCUDA}, and DS-CUDA~\cite{DSCUDA} realize GPU virtualization via API remoting without significant performance penalty.
To deal with software maintenance in high-level libraries, LoGV~\cite{LoGV}, GPUvm~\cite{GPUvm}, and gScale~\cite{gScale} implement Para and full virtualization at the GPU driver level.
Among the hardware virtualization methods~\cite{HardwareVirtualizationTrends}, Intel VT-d~\cite{IntelVTd} and AMD-Vi~\cite{AMDVi} fail to support sharing a single GPU between multiple VMs.
And NVIDIA GRID~\cite{NVIDIAGRID} overcomes the difficulties and supports multiplexing a single GPU between VMs.
However, the mechanism of GPU virtualization means that it cannot provide stronger compute capability than a single GPU server.

\subsection{Distributed Machine Learning}
As the era of big data comes, the scale of Machine Learning system is becoming larger in training data and model parameters.
Since increasing the compute capability of a single server costs massive systematic efforts~\cite{DistributedMachineLearningSurvey, diannao},
distributed Machine Learning is proposed to overcome these challenges.
However, such a distributed system requires more knowledge in parallel programming and multi-machine communication, 
forming bottlenecks for researchers.
Hence, programmers need to pay more attention to reducing the network latency and optimizing the whole system~\cite{lee2017speeding, Gaia}.

\subsection{Network Processor}
Network processors, like Intel IXP Series~\cite{intel_ixp}, are specialized integrated circuits (ICs) designed to efficiently handle and process network-related tasks, such as data packet routing, switching, security, and quality of service (QoS) management. 
They play a crucial role in enabling the high-speed data communication that underpins modern computer networks.
In contrast, \proxyname serves the purpose of converting between PCIe and network packets, as described in Section~\ref{sec:proxy}.
  
\section{Conclusion}
  
In this paper, we design a new implementation of \textit{datacenter-scale} \textit{GPU disaggregation} at the PCIe level, named \sysname.
\sysname is effective in \textit{full-scenario suppport}, \textit{software transparency}, \textit{datacenter-scale disaggregation}, \textit{large capacity}, and \textit{good performance}.
Meanwhile, in order to measure the performance overhead incurred by \sysname, we build up a performance model to make estimations.
Instructed by modeling results, we develop an implementation system of \sysname and conduct detailed experiments.
The experimental statistics demonstrate that the overhead of \sysname is acceptable in most cases, which is less than 10\%.
In addition, we make extra experiments to figure out causes of the performance overhead and propose optimization advice, which is regarded as our future work.

\section{Acknowledgements}

We thank the anonymous reviewers for their comments that greatly helped improve the presentation of this article.
The authors are partially
supported by the National Key R\&D Program of China (No.
2022YFE0113200), the National Natural Science Foundation
of China (NSFC) under Grant U21A20464, and Alibaba Group through Alibaba Research Intern Program. Any opinions, findings, conclusions, or recommendations expressed in
this material are those of the authors and do not necessarily
reflect the views of funding agencies.

\bibliographystyle{ACM-Reference-Format}
\bibliography{references}

\end{document}